\documentstyle[epsfig]{aipproc}

\def\ltap{\raisebox{-.4ex}{\rlap{$\sim$}} \raisebox{.4ex}{$<$}}
\def\rts{\sqrt s}
\def\wp{W^+}
\def\wm{W^-}
\def\anti{\overline}

\def\mev{~{\rm MeV}}
\def\gev{~{\rm GeV}}

\def\anti{\overline}

\def\eg{{\it e.g.}}

\def\br{B}

\def\h{h}
\def\mh{m_{\h}}

\def\mt{m_t}

\def\mw{M_W}
\def\mz{M_Z}

\def\lsim{\ltap}

\begin{document}
\hspace*{3.6in}{\bf IUHET-382}\\
\hspace*{3.6in}{\bf January 1998}
\title{Threshold Cross Section Measurements\thanks{Presented
at the Fourth International Conference on the
Physics Potential and Development of 
$\mu ^+\mu^-$ Colliders, December 10-12, 1997, San Francisco, CA.}}

\author{M. S. Berger}
\address{Indiana University\\
Bloomington Indiana 47405}

\maketitle

\begin{abstract}
Accurate measurements of particles masses, couplings and widths are possible
by measuring production cross sections near threshold. 
We discuss the prospects for performing such measurements at a high 
luminosity muon collider. 

\end{abstract}

\section*{Introduction}

A muon collider is particularly
well suited to the threshold measurement because the spread in energy of the 
beam is very small\cite{feas}.
Pair production of $W$-bosons, $t\bar{t}$ 
production and the Bjorken process $\mu^+\mu^-\to ZH$ have been considered as
possible places to study thresholds at a muon collider\cite{bbgh3,Zh,ttbar}.
Threshold production of chargino pairs at a muon collider 
offers a possible way of accurately measuring
the chargino mass\cite{prep}.

We assume here 
that the muon collider has a relatively modest beam energy spread of 
$R=0.1\%$, where $R$ is the rms spread of the energy of a muon 
beam.
We assume that 100~fb$^{-1}$ integrated luminosity is available and that
this amount of luminosity could be accumulated at 
the relevant energies for the measuring the threshold
cross sections; high
luminosity is essential if the threshold measurements are to prove
interesting. 

\section*{\boldmath $M_W$ Measurement at the $\mu^+\mu^-\to W^+W^-$ Threshold}

The threshold cross section is most
sensitive to $\mw$ just above $\sqrt s = 2\mw$,
but a tradeoff exists between
maximizing the signal rate and the sensitivity of the cross section to $\mw$.
Detailed analysis \cite{lepii} shows that
if the background level is small and systematic uncertainties in efficiencies
are not important, then the optimal measurement of $\mw$
is obtained by collecting data at a single energy
$$\sqrt{s} \sim 2\mw + 0.5  \gev\ \sim\ 161  \gev ,$$
where the threshold cross section is sharply rising.

At a muon
collider with high luminosity, systematic errors
arising from uncertainties in the background level and the detection/triggering
efficiencies will be dominant unless
some of the luminosity is devoted to measuring
the level of the background (which automatically includes somewhat similar
efficiencies) at an energy below the $W^+W^-$ threshold.
Then, assuming that efficiencies for the background and $\wp\wm$ signal are
sufficiently well understood that systematic uncertainties
effectively cancel in the ratio
of the above-threshold to the below-threshold rates, a very accurate $\mw$
determination becomes possible.

We analyzed\cite{bbgh3} the possible precision obtainable for the $W$ mass via
just two measurements: one at center of mass energy $\sqrt{s}=161\gev$, just
above threshold, and one at $\sqrt{s}=150\gev$. 
The optimal $M_W$ measurement is obtained by expending about two-thirds
of the luminosity at $\sqrt{s}=161\gev$ and  one-third
at $\sqrt{s}=150\gev$.
Combining the three modes, an overall precision of
$\Delta M_W = 6 \rm\ MeV$
should be achievable with 100~fb$^{-1}$ integrated luminosity.

\section*{\boldmath Higgs Boson Measurement at the
$\mu^+\mu^-\to Z\lowercase{h}$ Threshold}

The SM Higgs boson is easily discovered in the Bjorken Higgs-strahlung
process~\cite{bj} $\ell^+\ell^-\to Z\h$
running the machine well above threshold, \eg\ at $\rts=500\gev$.
For $\mh\lsim 2\mw$ the dominant Higgs boson decay is to $b\anti b$
and most backgrounds can be eliminated by $b$-tagging.
A very accurate determination of $\mh$ could then obtained
by measuring the threshold cross section of $Z\h$ production, which rises 
rapidly as shown in Fig.~1(a) since the threshold
behavior is $S$-wave.

\begin{figure}[htb]
\leavevmode
\begin{center}
\epsfxsize=2.50in\hspace{0in}\epsffile{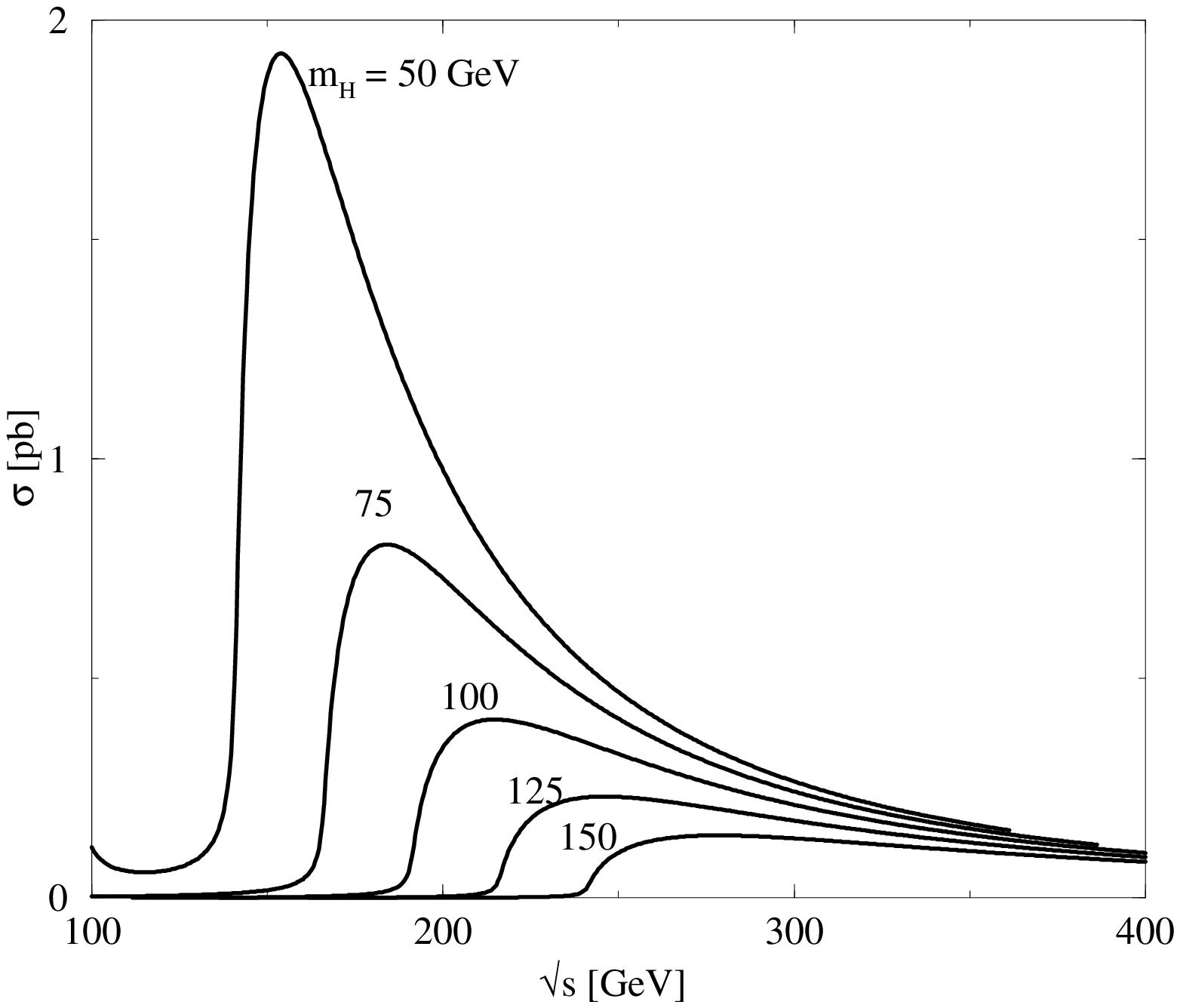}
\epsfxsize=2.50in\hspace{0.5in}\epsffile{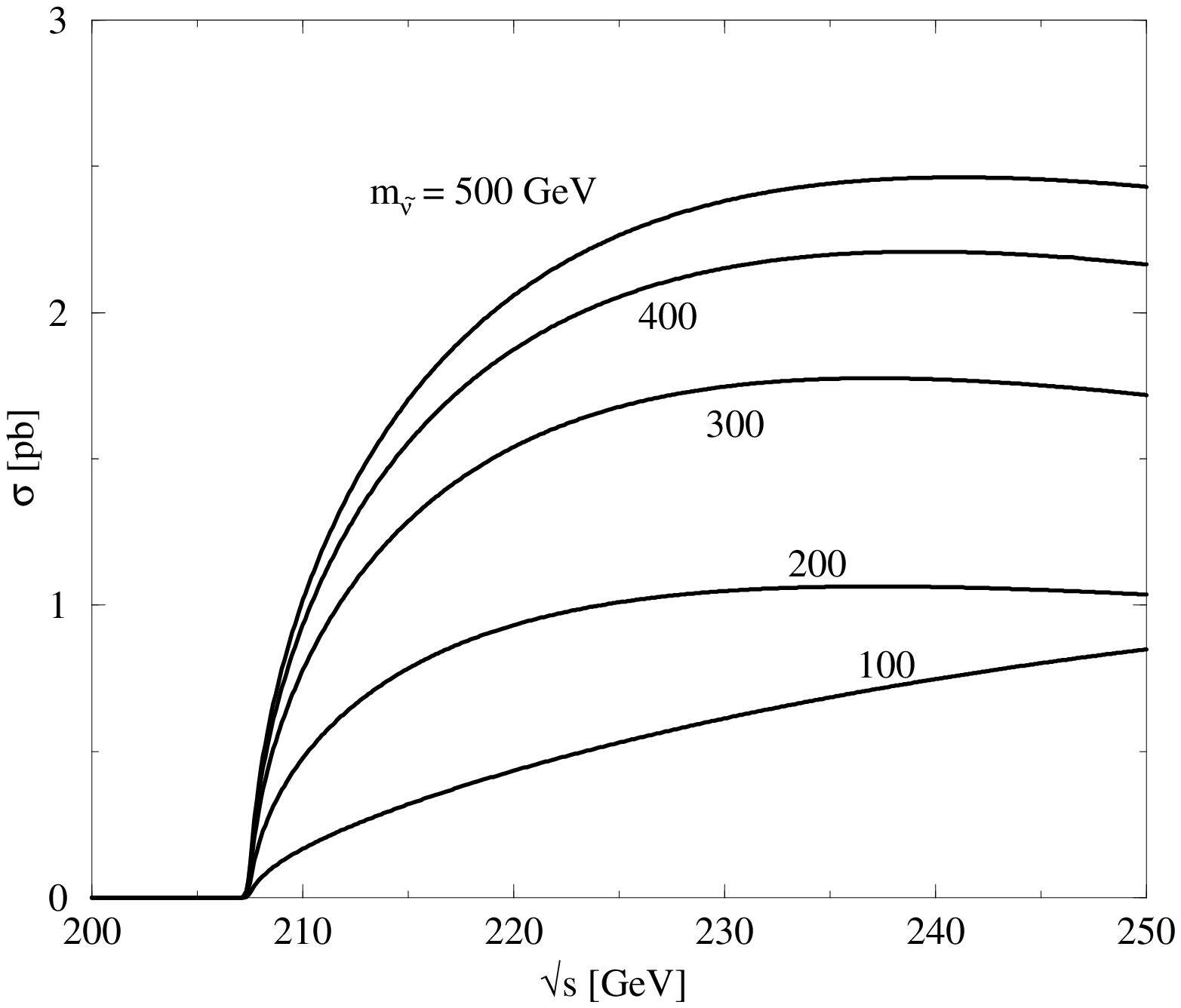}
\end{center}
\caption[]{\footnotesize\sf The cross section vs. $\protect\rts$
for (a) the process
$\mu^+\mu^-\to Z^\star\h\to f\anti f\h$ for a range of Higgs masses, and 
for (b)
$\mu^+\mu^-\to \tilde{\chi}^+\tilde{\chi}^-$ for various sneutrino
masses and $m_{\tilde{\chi}^{\pm}}=103.7$~GeV. }
\label{zhcurve}
\end{figure}

The sensitivity to the SM Higgs boson mass is maximized by
a single measurement of the cross section at $\rts=\mz+\mh+0.5$~GeV,
just above the real particle threshold provided that the normalization of 
the measured $Z\h$ cross section
as a function of $\rts$ can be precisely predicted, including
efficiencies and systematic effects. 
We employed $b$-tagging and cuts in order to
reduce the background to a very low level. These cuts and other systematic
uncertainties are discussed in more detail in Ref.~\cite{Zh}.
The background is very much smaller than the signal unless $\mh$
is close to $\mz$. The electroweak radiative corrections to the 
cross section are
estimated to be less than 1\% for $m_H\sim100$~GeV\cite{kniehl}, and the 
measurement of the cross section described here is at the 2\% level.
We found a  precision of the SM Higgs mass determination to within 45\mev\
for $\mh=100\gev$ may be achievable at a muon collider.
More generally the precision ranges from 20-100~MeV for $\mh < 150$~GeV.

Beyond the Standard Model the cross section generally depends on 
the $ZZH$ coupling ($g_{ZZ\h}^{}$) and the total Higgs width 
($\Gamma_H$) in addition to $\mh$.
In order to simultaneously determine these three quantities,
measurements could be made at the three c.m.\ energies
$\rts=\mh+\mz+20\gev$,
$\rts=\mh+\mz+0.5\gev$, and
$\rts=\mh+\mz-2\gev$.
With a three-parameter fit to 
$\mh$, $g_{ZZ\h}^2\br(\h\to b\anti b)$ and $\Gamma_H$,
the attainable error in $\mh$ is about $110\mev$ at the $1\sigma$ level for 
a 100~GeV Higgs.
Measurements that would simultaneously determine $\mh$,
$\sigma(Z\h)\br(\h\to b\anti b)$ and $\Gamma_H^{}$ could be done at
a level of accuracy that could distinguish a Standard Model
Higgs boson from its many possible (\eg\ supersymmetric) extensions\cite{Zh}.

\section*{\boldmath Top-quark Mass Measurement at the
$\mu^+\mu^-\to \lowercase{t \bar t}$ Threshold}

The top-quark threshold cross section is
calculable since the large
top-quark mass puts one in the perturbative regime of QCD\cite{fk}.
One can perform scan of the
threshold curve by devoting to 10~fb$^{-1}$ integrated luminosity
to measuring the cross section at each of ten energies in
1~GeV intervals. Then the top-quark mass
can be determined to within $\Delta\mt\sim 70\mev$,
provided systematics and theoretical uncertainties are under control. 
Considerable progress has been made recently in the theoretical 
calculations of the some NNLO corrections to the threshold cross 
section\cite{hoang}. The remaining theoretical uncertainties\cite{jezabek} 
in the threshold cross section are still fairly large and make it difficult
to fully exploit the large luminosity for determining say the strong
coupling $\alpha_s$ or a light Higgs boson mass (and the top quark 
Yukawa coupling) from the size of 
the cross section.
Furthermore there is theoretical ambiguity in the
mass definition of the top quark. The 
theoretical ambiguity in relating quark pole mass to other definitions of
the top quark mass (that might be relevant as input to radiative correction
calculations) is of  order
$\Lambda_{QCD}$, {\it i.e.}, or a few hundred MeV \cite{sw}. So it is not
clear that an extraction of the top-quark mass better than this is useful, at 
least at the present time.

\section*{Chargino Signal and Background}

The mass of the 
lighter chargino in the minimal supersymmetric
standard model (MSSM) can be determined accurately by measuring the 
cross section\footnote{The measurement
of the chargino mass via the threshold cross section has been considered 
previously for electron-positron machines in Ref.~\cite{oldthresh,nlcstudy}.
We consider the measurement at a muon collider with 
high luminosity, carefully taking into account the beam effects and 
reoptimizing cuts to eliminate the background in the threshold region.}
for 
\begin{equation}
\mu^+\mu^- \to \chi^+ \chi^-
\end{equation}
near the threshold\cite{prep}.
The precision that can be obtained in the chargino mass depends substantially
on the mass of the chargino mass itself: the heavier the chargino the smaller
the production cross section. The cross section also depends on the mass of the
sneutrino which appears in the $t$-channel since this contribution
interferes destructively with the $s$-channel graphs. The cross section is 
displayed in Fig.~1(b) for several values of the sneutrino mass.
If the 
lightest chargino is gaugino-dominated, then changing the parameters of the 
chargino mass matrix essentially changes the mass but not the chargino 
couplings significantly. The width of the lightest chargino is usually less
than a few MeV, and often substantially less when two-body decays are 
kinematically impossible.
Therefore one can envision a measurement of the 
cross section
that depends on just two parameters: the chargino mass $m_{\tilde{\chi}^{\pm}}$
and the sneutrino mass $m_{\tilde{\nu}}$.\footnote{The overall normalization
of the cross section could also depend on radiative corrections which 
could be substantial in some cases\cite{rc}.}

As in the other threshold
measurements, the statistical precision on the chargino mass is maximized 
just above $2m_{\tilde{\chi}^{\pm}}$. A simultaneous measurement of the 
chargino and sneutrino masses requires a 
sampling of the cross section at at least two points. 
It turns out to be advantageous for the chargino mass measurement to choose 
this higher energy measurement at a point where the chargino cross section 
is not flat.

The chargino decay mode is 
$\tilde{\chi}^{\pm}\to \tilde{\chi}^0f\overline{f}^\prime$ provided 
the chargino
is lighter than the muon sneutrino. 
The cross section is reduced near threshold, so 
the cuts to reduce backgrounds need to be reoptimized. 
The backgrounds to chargino pair-production have been investigated in 
Refs.~\cite{tfmyo,grivaz} where the signal efficiencies have been obtained 
for the various final states when the center-of-mass energy is 
$\sqrt{s}=500$~GeV. The primary background is $W$ pair production which is 
very large, but can be effectively eliminated because the $W$'s are produced 
in the very-forward direction. However, if the energy is reduced so that the
collider is operating in the chargino threshold region, then the effectiveness
of these cuts might be reduced (the signal events might be expected to be 
more spherical as well). Therefore the efficiencies were reinvestigated
for the threshold measurement.

A further advantage of the threshold measurement is that the chargino mass
measurement is somewhat isolated from its subsequent decays. Distributions in
the final state observables, say e.g. $E_{jj}$ from the decay
$\tilde{\chi}^{\pm}\to \tilde{\chi}^0jj$\cite{tfmyo}, depend on the 
neutralino mass. The cross section for chargino pair production, on the 
other hand, is independent of the final state particles, and only the 
branching fractions and detector efficiencies for the various final states
impact this measurement (as indicated above,
if $m_{\tilde{\chi}^{\pm}}-m_{\tilde{\chi}^0}>M_W$
the branching fractions of chargino decay is given essentially in terms of 
the $W$ branching fractions).

The chargino production cross section decreases with increasing chargino mass.
Therefore the precision with which the mass can be measured is better at 
smaller values of the mass with precisions of as small as 30~MeV possible
for $m_{\tilde{\chi}^{\pm}}=100$~GeV. For $m_{\tilde{\chi}^{\pm}}=200$~GeV 
the chargino mass can be determined
to 100 (200)~Mev for $m_{\tilde{\nu}}=500$ (300)~GeV.
The sneutrino mass can be measured to about 6~GeV accuracy for 
$m_{\tilde{\nu}} = 300$~GeV and to about 20~GeV accuracy for 
$m_{\tilde{\nu}} = 500$~GeV. This provides an indirect method of 
measuring the sneutrino mass (the sneutrino 
might be too heavy to produce directly).

\section*{Conclusion}

A muon collider would provide an opportunity for precision 
mass measurements in the respective threshold 
regions\footnote{The most recent TESLA design
envisions a beam energy spread of $R=0.2\%$\cite{miller} while the
NLC design expects a beam energy spread of $R=1.0\%$. A high
energy $e^+e^-$ collider in the large VLHC tunnel would have a beam 
spread of $\sigma _E=0.26$~GeV\cite{norem} which should give numbers
precisions comparable to those considered here.}.
The precisions that can be obtained for particle masses is shown
in Table~1 assuming an integrated luminosity of 100~${\rm fb}^{-1}$.
The precisions for the Higgs and chargino measurements are correlated
with the (as of yet unknown) mass, so the ranges we considered are shown
as well.
To utilize the highest precision measurements achievable at the statistical
level, theoretical uncertainties
and other systematics need to be under control in all cases. 
The muon sneutrino mass 
can also be simultaneously measured to a few GeV if it is less than 500~GeV
in the process $\mu^+\mu^- \to \chi^+ \chi^-$.

\begin{table}
\caption{Precison of mass measurements assuming 100~fb$^{-1}$ luminosity.
The ranges considered for the Higgs and chargino masses are also shown. }
\label{table1}
\begin{tabular}{ldd}
   Particle & Mass Measurement (MeV) & Mass Range (GeV) \\
\tableline
$W$ & 6 & -- \\
$t$ & 70 & -- \\
$\h$ & 20-150  & 50-200 \\
$\chi^\pm$ & 30-200 & 100-200\tablenote{$m_{\tilde{\nu}}>300$~GeV}
\end{tabular}
\end{table}

\section*{Acknowledgments}

I thank V.~Barger, J.~F.~Gunion and T.~Han for a pleasant collaboration
on the issues reported here.
This work was supported in part by the U.S. Department of Energy
under Grant
No.~DE-FG02-91ER40661.


\begin{references}

\bibitem{feas} {\it $\mu^+\mu^-$ Collider: A Feasibility Study},
Snowmass, Colorado, July, 1996.

\bibitem{bbgh3} V.~Barger, M.S.~Berger, J.F.~Gunion and T.~Han,
Phys.\ Rev.\ {\bf D56}, 1714 (1997).

\bibitem{Zh} V. Barger, M. S. Berger, J. F. Gunion and T. Han,
Phys.\ Rev.\ Lett.\ {\bf 78}, 3991 (1997).

\bibitem{ttbar}
M.S.~Berger, talk presented at the
{\it Workshop on Particle Theory and Phenomenology:
Physics of the Top Quark}, Iowa State University,
May 25--26, 1995, hep-ph/9508209.

\bibitem{prep} V.~Barger, M.S.~Berger and T.~Han, hep-ph/9801410.

\bibitem{lepii} Z. Kunszt and W.J. Stirling {\it et al.},
hep-ph/9602352,
in {\it Proceedings of the Workshop on Physics at LEP2}, eds.\
G. Alterelli, T. Sjostrand and F. Zwirner, CERN Yellow Report
CERN-96-01 (1996), Vol.~1, p.~141; W.J. Stirling, Nucl. Phys.
{\bf B456}, 3 (1995).

\bibitem{bj} J.D.~Bjorken, {\it Proceedings of the Summer Institute on
Particle Physics}, ed. M.~Zipf (Stanford, 1976).

\bibitem{kniehl} B.A. Kniehl, Z. Phys.\ {\bf C55}, 605 (1992);
R. Hempfling and B. Kniehl, Z.~Phys.\ {\bf C59}, 263 (1993).

\bibitem{fk} V.S.~Fadin and V.A.~Khoze, JETP Lett.\ {\bf  46}, 525 (1987);
Sov. J. Nucl. Phys. {\bf  48}, 309 (1988).

\bibitem{hoang} A.~H.~Hoang, Phys.\ Rev. {\bf D56},
5851 (1997); A.~H.~Hoang, Talk at the Workshop on Physics at the
First Muon Collider and at the Front End of the Muon Collider, Batavia, IL, 
6-9 Nov 1997, hep-ph/9801273; A.~H.~Hoang and T.~Teubner, hep-ph/9801397.

\bibitem{jezabek} M.~Jezabek, et al., hep-ph/9801419.

\bibitem{sw} M.~C.~Smith and S.~Willenbrock, Phys.\ Rev.\ Lett.\ {\bf 79}, 
3825 (1997). 

\bibitem{oldthresh} A.~Leike, Int.\ J.\ Mod.\ Phys.\ {\bf A3}, 2895 (1988).

\bibitem{nlcstudy} {\it Physics with $e^+e^-$ Linear Colliders},
by ECFA/DESY LC Physics Working Group (E. Accomando et al.), DESY-97-100, 
May 1997, hep-ph/9705442. 

\bibitem{rc} P.~Chankowski, Phys.\ Rev.\ {\bf D41}, 2877 (1990); 
M.~M.~Nojiri, K.~Fujii and T.~Tsukamoto, Phys.\ Rev.\ {\bf D54}, 6756 (1996);
H.-C.~Cheng, J.~L.~Feng and N.~Polonsky, Phys.\ Rev.\ {\bf D56}, 6875 (1997);
Phys.\ Rev.\ {\bf D57}, 152 (1998);
M.~A.~Diaz, S.~F.~King and D.~A.~Ross, hep-ph/9711307.

\bibitem{tfmyo} T.~Tsukamoto, K.~Fujii, H.~Murayama, M.~Yamaguchi and 
Y.~Okada, Phys.\ Rev.\ {\bf D51}, 3153 (1995).

\bibitem{grivaz} J.-F.~Grivaz, preprint LAL 91-63, Talk at the Workshop on 
Physics and Experiments with Linear Colliders, Saariselka, Finland, 
9-14 September 1991.

\bibitem{miller} D.~Miller, private communication.

\bibitem{norem} J.~Norem, private communication and\\ 
http://www-ap.fnal.gov/VLHC/electrons/index.html.

\end{references}
\end{document}